\documentclass{ws-procs975x65}

\usepackage{amsmath}
\usepackage{amsfonts}
\usepackage{amssymb}
\usepackage{cite}

\DeclareMathOperator{\extdm}{d}
\newcommand{\extd}{\extdm \!}

\newcommand{\oB}{\vert_{\partial\mathcal{M}}}

\newcommand{\half}[1]{\ensuremath{\frac{#1}{2}}}

\newcommand{\varfrac}[2][]{\frac{\delta #1}{\delta #2}}

\begin{document}



\title{BLACK HOLES AS BOUNDARIES \\IN 2D DILATON SUPERGRAVITY}

\author{LUZI BERGAMIN}

\address{ESA Advanced Concepts Team, ESTEC -- DG-PI\\ Keplerlaan 1, 2201 AZ Noordwijk, The Netherlands\\
\email{Luzi.Bergamin@esa.int}}

\author{DANIEL GRUMILLER}

\address{Center for Theoretical Physics,
Massachusetts Institute of Technology\\
77 Massachusetts Ave.,
Cambridge, MA 02139, USA\\
\email{grumil@lns.mit.edu}}


\begin{abstract}
We discuss 2D dilaton supergravity in the presence of boundaries. Generic ones lead to results different from black hole horizon boundaries. In particular, the respective numbers of physical degrees of freedom differ, thus generalizing the bosonic results of {\tt hep-th/0512230}.
\end{abstract}

\bodymatter

\section{Introduction}
Frequently it is argued that the microstates responsible for the Bekenstein-Hawking entropy should arise from some physical degrees of freedom located near or on the black hole horizon (cf.~e.g. Ref.~\refcite{Carlip:2006fm} and references therein). Recently we have provided evidence within the framework of 2D dilaton gravity that instead entropy may emerge from the conversion of physical degrees of freedom, attached to a generic boundary, into unobservable gauge degrees of freedom attached to the horizon \cite{Bergamin:2005pg,Bergamin:2006zy}. In this joint proceedings contribution we generalize such considerations to 2D dilaton supergravity (SUGRA).

We start with the first order 2D dilaton SUGRA action\footnote{The superspace action by Park and Strominger \cite{Park:1993sd} describes the same theory and has several advantages over \eqref{1.1}. However, the solution of all constraints, the construction of classical solutions and path integral quantization is much simpler starting with the first order action.}
\begin{equation}
\label{1.1}
S = \int_{\mathcal M} X^I \extd A_I + \half{1} P^{IJ} A_J \wedge A_I\,.
\end{equation}
We use a notation that is a convenient mixture between the one employed in our previous paper on the subject\cite{Bergamin:2005pg} (consistent with Ref.~\refcite{Grumiller:2002nm}) and our papers on SUGRA \cite{Bergamin:2002ju,Bergamin:2003am,Bergamin:2003mh,Bergamin:2004us}.
The graded 1-form fields $A_I$ comprise the (dual) spin-connection $\omega$, the Zweibeine $e_{\pm\pm}$ and the gravitino $\psi_\pm$. The graded 0-form fields $X^I$ comprise the dilaton $\phi$, Lagrange-multipliers for torsion $X^{\pm\pm}$ and the dilatino $\chi^\pm$. They span a target-space equipped with a Poisson tensor $P^{IJ}$, viz., a (graded) Poisson manifold \cite{Schaller:1994es}. The Poisson tensor is given by Eqs.~(2.8), (2.16)-(2.19) in Ref.~\refcite{Bergamin:2004us}; we refrain from presenting these formulas here. The action \eqref{1.1} is \emph{not} consistent with the Gibbons-Hawking-York prescription used in Ref.~\refcite{Bergamin:2005pg} but nevertheless a valid (and for various purposes useful) starting point. It is advantageous to define the canonical variables
\begin{align}
q_I &= A_{1I}\ , & p^I &= X^I\ , & \bar q_I &= A_{0I}\ , & \bar p^I &\approx 0\ .
\end{align}
The indices $0,1$ refer to world-sheet coordinates $x^0,x^1$, where $x^0$ plays the role of time. Evidently, the canonical momenta $\bar p^I$ are primary constraints. We keep all conventions regarding spinor calculus as defined e.g.~in the appendix of Ref.~\refcite{Bergamin:2003am}. 
To keep a simple representation of the constraints we need a Poisson bracket with
\begin{equation}
\{ q_I , p'^J \} = (-1)^{I\cdot J+1} \{p'^J, q_I\} = \delta^J_I \delta(x-x')\ ,
\end{equation}
which can be achieved by the definition
\begin{equation}
\label{1.4}
  \begin{split}
    \{A, B'\} &= \int_{x''} \Bigl[\bigl( (-1)^{A \cdot I + 1} \varfrac[A]{{p''}^I}
    \varfrac[B']{q''_I} + (-1)^{I(A+1)} \varfrac[A]{q''_I}
    \varfrac[B']{{p''}^I}\bigr) + (q\rightarrow \bar{q}, p \rightarrow
    \bar{p}) \Bigr] 
  \end{split}
\end{equation}
The boundary $\partial\mathcal{M}$ is supposed to be a hypersurface of constant $x^1$. As in \cite{Bergamin:2005pg} it is considered to be a \emph{lower} one.

In Section \ref{se:2} we present results of the constraint analysis and possible choices for boundary conditions. In Section \ref{se:3} we discuss the gauge fixing and construct the reduced phase space. The interpretation of our results is analogous to the bosonic one in Refs.~\refcite{Bergamin:2005pg,Bergamin:2006zy}, so we focus on issues peculiar to SUGRA.

\section{Constraint analysis and boundary conditions}\label{se:2}

With standard methods we obtain the secondary constraints
\begin{equation}
G^I[\eta] = \int \extd x^1 \left( \partial_1 p^I + P^{IJ} q_J \right) \eta + p^I \eta \oB \approx 0\,.
\end{equation}
The constraint algebra including boundary terms reads
\begin{equation}
\label{1.6}
\{G^I[\eta],G^J[\xi]\} = G^K[\eta\xi] \partial_K P^{IJ} - \left( p^K \partial_K - 1 \right) P^{IJ} \eta \xi\oB\,.
\end{equation}
Notice that all brackets $\{\bar p^I, G^J \}$ vanish with this choice of the boundary action in contrast to Ref.~\refcite{Bergamin:2005pg}. Moreover, the boundary term in \eqref{1.6} vanishes whenever the Poisson tensor is homogeneous of degree one. This is always true for the generators of local Lorentz transformations, i.e.\ for the brackets $\{G^\phi[\eta],G^I[\xi]\}$, and the basic relation defining the supersymmetry algebra $\{G^\pm[\eta], G^\pm[\xi] \} = -2 \sqrt{2} G^{\pm\pm}$. Among the purely bosonic models the boundary terms vanish completely for the Jackiw-Teitelboim model \cite{jackiw:1984}, for the Witten black hole \cite{Witten:1991yr} and for models with an $(A)dS_2$ ground state, as noted in Ref.~\refcite{Grumiller:2006rc}. 
This characteristic is retained upon supersymmetrization because the full Poisson tensor is homogeneous of degree one if the bosonic sector exhibits this property.

Variation of the action \eqref{1.1} yields the boundary conditions
\begin{equation}
\label{2.1}
 p^I \delta \bar q_{I}\oB = 0\ . 
\end{equation}
As in Ref.~\refcite{Bergamin:2005pg} we implement them by means of constraints on the phase space with support at the boundary only.
The choices for the three bosonic components are similar to that work and will be recapitulated briefly below.  Here we concentrate on the fermionic variables, where two different choices of boundary constraints,
\begin{align}
\label{2.2.1}
 B_\pm[\eta] &= (\bar{q}_\pm - \mathcal A_{\pm}) \eta\oB && \text{or} & \hat B^\pm[\eta] &= p^\pm \eta\oB\ ,
\end{align}
exist. A mixture of the two for the different components of the spinors is conceivable. 

To see how the different choices can affect the result one has to construct the line element [cf.\ eqs.\ (100) and (101) in Ref.~\refcite{Bergamin:2003am}]. 
Not surprisingly, all fermionic contributions to the line element vanish at the boundary if both components of the dilatino are set to zero there. But even with one component of the dilatino set to zero the bosonic result for the Killing norm emerges, as the (classical) space of anti-commuting variables is too small to contribute to a bosonic quantity. If instead of the dilatino both components of the gravitino are fixed at the boundary, $p^{++}p^{--}$ need no longer be proportional to the Killing norm (this conclusion does not depend on the value of the gravitino chosen at the boundary.) We do not go into further details of this question here, but simply stick to the first two choices of boundary conditions, i.e., we always fix at least one dilatino component at the boundary.

The bulk theory contains only first class constraints. However,
due to possible boundary contributions in \eqref{1.6} and as a consequence of
the boundary constraints enforcing \eqref{2.1}, terms are generated in the
evaluation of Poisson brackets with support exclusively at
the boundary. They convert some of the constraints into second class.
This feature was observed already in the bosonic case \cite{Bergamin:2005pg,Bergamin:2006zy}. We shall discuss now its extension to SUGRA.

\paragraph{Generic Boundary}
For a generic boundary to solve the boundary problem \eqref{2.1} among the bosonic variables the only possible choice is $\delta \bar q_{I} = 0$, which we implement by means of the constraints
\begin{equation}
\label{2.2}
B_i[\eta] = (\bar{q}_i - \mathcal A_{i}) \eta\oB\ .
\end{equation}
The only constraint that remains first class for all possible choices in \eqref{2.2.1} is the Lorentz constraint $G^\phi$. Besides $G^\phi$ there can remain up to two components of $\bar p^\pm$ first class depending on the choice in \eqref{2.2.1}. The remaining secondary constraints become second class due to boundary contributions in \eqref{1.6} and possibly additional contributions from brackets with $\hat B^\pm$. 
Moreover, because of
\begin{equation}
\label{2.3}
\{B_I[\eta],\bar{p}^J[\xi]\} = \delta_I^J \eta \xi\oB
\end{equation}
the $B_I$ make the primary constraints second class. 


\paragraph{Horizon}
As motivated in Ref.~\refcite{Bergamin:2005pg} a horizon is best described by 
\begin{equation}
\label{2.4}
\delta \bar q_\phi\oB = \delta \bar q_{++}\oB = p^{--}\oB = 0\ .
\end{equation}
Consistency with the equations of motion implies $\bar q_{++}\oB =0$ as well. Inspecting the general solution of the SUGRA model (cf. section 6 of Ref.~\refcite{Bergamin:2003am}) it appears to be self-evident to choose $p^+ = p^- = 0$ as boundary conditions of the fermionic sector. However, it should be noticed that this is not enforced.

Again all secondary constraints except the Lorentz constraint become second class due to the boundary terms in \eqref{1.6} (some of the contributions vanish weakly due to the $\hat B^I$ constraints, but this is not sufficient to keep an additional constraint first class.) Among the primary constraints $\bar p^{--}$ and $\bar p^\pm$ remain first class while all $B_I$ and $\hat B^I$ become second class. For consistency it is then seen that a linear combination of the second class constraints actually remains first class (the ``Dirac matrix'' has determinant zero.)

\vspace{2ex}
In summary a difference between a generic boundary and a horizon is seen at the level of the constraint algebra similar to the result of Ref.~\refcite{Bergamin:2005pg}: if the boundary is a horizon more first class constraints are present than in the case of a generic horizon. Therefore, if the boundary is a horizon there are more gauge degrees of freedom and fewer physical degrees of freedom.

As mentioned above the boundary terms in \eqref{1.6} vanish for a certain class of models, in which case more first class constraints are encountered. Notice that some of the $G^I$ still turn into second class constraints due to the Poisson brackets with $\hat{B}^I$ from \eqref{2.2.1} and/or \eqref{2.4}.

\section{Gauge fixing and reduced phase space}\label{se:3}

In order to exhibit explicitly the conversion of physical into gauge degrees of freedom we now construct the reduced phase space in analogy to Ref.~\refcite{Bergamin:2005pg}. In case of a generic boundary the gauge\footnote{Notice that according to our conventions the light-cone components of a vector are purely imaginary \cite{Bergamin:2003am,Bergamin:2004us}.} $q_{++} = -i$ and $q_I=0$ for all other $I$ can be used, yielding the straightforward result:
\begin{gather}
 \label{3.1}
\begin{alignat}{2}
 G^{++}:&\quad p^{++} = \hat p^{++}(x^0)\ , &\qquad G^\phi:&\quad p^\phi = \hat p^\phi(x^0) + i x^1 \hat p^{++}\ ,\\
 G^+:&\quad p^+= \hat p^+(x^0)\ , &\qquad G^-:&\quad p^- = e^{-\frac{Q}{2}}\left( \hat p^-(x^0) + \sqrt{2} \frac{\hat p^+(x^0)}{\hat p^{++}(x^0)} w \right)\ ,
\end{alignat}\\
G^{--}:\  p^{--}= \frac{e^{-Q}}{\hat p^{++}(x^0)} \left( \hat p^{--}(x^0) - W - \frac{1}{2} \frac{\hat p^-(x^0) \hat p^+(x^0)}{\hat p^{++}(x^0)} w' \right)\ .\label{3.3}
\end{gather}
Here $Q$, $W$ and $w$ are all functions of the dilaton $p^\phi=\phi$, cf.~Ref.~\refcite{Bergamin:2003am} for their definitions. At this point it matters which boundary conditions were chosen. Quite generally each choice of a boundary constraint $\hat B^I$ fixes one of the free functions in \eqref{3.1}-\eqref{3.3}, as the analytic continuation of the bulk solution to the boundary must be equivalent to the boundary value. In the fermionic sector this means that boundary degrees of freedom can be present only if we fix the gravitino at the boundary. This conclusion is independent of the nature of the boundary (generic boundary vs.\ horizon.)

To proceed it is important to define the boundary conditions in the fermionic sector. If we choose $p^+\oB=p^-\oB=0$ all fermionic integration constants in \eqref{3.1}-\eqref{3.3} are removed. The derivation and the results within the bosonic sector are the same as in the purely bosonic case, since in all relevant equations explicit fermionic contributions are set to zero by means of the boundary conditions. Like in Ref.~\refcite{Bergamin:2005pg} the gauge fixing procedure changes if a generic boundary is replaced by a horizon. Notice however, that the gauge used in Ref.~\refcite{Bergamin:2005pg} [cf.\ eq.\ (6.10) therein] is not suitable here, as it would fix the boundary value of the dilaton which remains free in the current approach. A possible choice is to replace $q_\phi = 0$ by $p^{++} = i$, which together with the boundary constraint $p^{--}$ removes two bosonic degrees of freedom.
There remains the possibility to fix one component of the gravitino and one of the dilatino. This turns out to be an especially interesting case as one finds that the boundary prescription for a horizon
\begin{align}
 \delta \bar q_\phi\oB &= \delta \bar q_{++}\oB = \delta \bar q_{+}\oB =0 &  p^{--}\oB &= p^-\oB = 0 \label{eq:1783}
\end{align}
together with the equations of motion implies not just ${\bar q_{++}}\oB = 0$ but also ${\bar q_{+}}\oB = 0$. Then it can be checked that this leaves two symmetry parameters ($\epsilon_{--}$ and $\epsilon_-$ in the notation of Ref.~\refcite{Bergamin:2003mh}) unrestricted at the boundary. The algebra closes trivially among the unbroken symmetries as all commutators vanish identically. This implies the necessity of yet another gauge condition. A possible choice is
\begin{align}
\label{4.1}
 q_{++} &= -i\ , & q_{--} &= 0\ , & p^{++} &= i\ , & q_+ &= 0\ , & p^+ &= 0\ .
\end{align}
This eliminates two bosonic boundary degrees of freedom at the horizon, but only one fermionic one because one can choose $p^-=0$ as boundary condition in the generic case as well. Thus, the phenomenon of phase-space reduction through horizon constraints readily generalizes from the purely bosonic case \cite{Bergamin:2005pg,Bergamin:2006zy} to SUGRA. 

The existence of unbroken supersymmetries at the boundary is not necessarily connected to the existence of BPS states. In the present case, however, it is easily seen that the ground state of a horizon respecting half of the supersymmetries actually is a BPS state. For solutions with vanishing fermions the only condition for a BPS state is a vanishing body of the Casimir function (mass)\cite{Bergamin:2003mh}
\begin{equation}
 \label{4.3}
M = 2 w^2 - p^{++} p^{--} e^Q = 0\,.
\end{equation}
A BPS solution therefore requires $w(\phi)\oB = 0$. Due to the quadratic nature of the first term in \eqref{4.3} it is obvious that the mass attains its minimum in the case of a BPS state and in this sense the latter is the ground state.
Once the gauge \eqref{4.1} is chosen it is easy to see that all classical solutions have vanishing fermions. Therefore, in this particular gauge \emph{all} states with $C=0$ actually are ground states.

It is worthwhile pointing out that the boundary conditions \eqref{eq:1783} are quite different to the ones in Ref.~\refcite{vanNieuwenhuizen:2005kg}. First we use a different boundary action than therein and second we choose as boundary a horizon. Even with the alternative prescription \`a la Gibbons-Hawking-York it is easy to show that a supersymmetric solution of the variational principle for a horizon is (again) quite different from the one for a generic boundary. In the latter case one has to choose a vanishing trace of the extrinsic curvature, in the former this clearly is not an option as the extrinsic curvature is not even well defined.

Finally, we would like to comment on the duality presented
recently \cite{Grumiller:2006xz}, which connects two different actions
\eqref{1.1} leading to the same set of classical solutions for the
line element. It was established at the classical level, without
supersymmetry and in the absence of boundaries, only. It is of
interest to check what happens when boundaries and supersymmetry are
included. As the boundary terms are
insensitive to the choice of the potentials an extension of the duality to
the case with boundaries is straightforward. Besides redefining the potentials the duality exchanges the
constant of motion with a dimensionful coupling constant in the
action. For bosonic models allowing a SUGRA extension both of their
signs are restricted.  The duality maps the positive coupling/positive
mass sector of the original theory to the negative coupling/negative
mass sector of the dual theory. Thus, the physical sector of the original (dual) model is mapped to the unphysical sector of the dual (original) model.

\section*{Acknowledgments}

We are grateful to Wolfgang Kummer and Dimitri Vassilevich for collaboration on Ref.~\refcite{Bergamin:2005pg}. 

This work is supported in part by funds provided by the U.S. Department of Energy (DOE) under the cooperative research agreement DEFG02-05ER41360.
DG has been supported by the Marie Curie Fellowship MC-OIF 021421 of the European Commission under the Sixth EU Framework Programme for Research and Technological Development (FP6).


\begin{thebibliography}{10}

\bibitem{Carlip:2006fm}
S.~Carlip, {\em Horizons, constraints, and black hole entropy}, {\tt hep-th/0601041}.

\bibitem{Bergamin:2005pg}
L.~Bergamin et.~al., 
{\em Class. Quant. Grav.} {\bf 23}, 3075 (2006).

\bibitem{Bergamin:2006zy}
L.~Bergamin and D.~Grumiller, {\em Killing horizons kill horizon degrees}, {\tt gr-qc/0605148}.

\bibitem{Park:1993sd}
Y.-C. Park and A.~Strominger, {\em Phys. Rev.} {\bf D47}, 1569 (1993).

\bibitem{Grumiller:2002nm}
D.~Grumiller, W.~Kummer and D.~V. Vassilevich, {\em Phys. Rept.} {\bf 369}, p.
  327 (2002).

\bibitem{Bergamin:2002ju}
L.~Bergamin and W.~Kummer, {\em JHEP} {\bf 05}, p. 074 (2003).

\bibitem{Bergamin:2003am}
L.~Bergamin and W.~Kummer, {\em Phys. Rev.} {\bf D68}, p. 104005 (2003).

\bibitem{Bergamin:2003mh}
L.~Bergamin, D.~Grumiller and W.~Kummer, {\em J. Phys.} {\bf A37}, 3881 (2004).

\bibitem{Bergamin:2004us}
L.~Bergamin, D.~Grumiller and W.~Kummer, {\em JHEP} {\bf 05}, p. 060 (2004).

\bibitem{Schaller:1994es}
P.~Schaller and T.~Strobl, {\em Mod. Phys. Lett.} {\bf A9}, 3129 (1994).

\bibitem{jackiw:1984}
R.~Jackiw and C.~Teitelboim, in {\em
  Quantum theory of gravity: Essays in honor of the 60th birthday of Bryce
  S.DeWitt\/},  ed. S.~Christensen (Hilger, Bristol, 1984).

\bibitem{Witten:1991yr}
E.~Witten, {\em Phys. Rev.} {\bf D44}, 314 (1991).
\item[]G.~Mandal, A.~M. Sengupta and S.~R. Wadia, {\em Mod. Phys. Lett.} {\bf A6}, 1685 (1991).
\item[]S.~Elitzur, A.~Forge and E.~Rabinovici, {\em Nucl. Phys.} {\bf B359}, 581 (1991).

\bibitem{Grumiller:2006rc}
D.~Grumiller and R.~Meyer, {\em Ramifications of lineland}, {\tt hep-th/0604049}.

\bibitem{vanNieuwenhuizen:2005kg}
P.~van Nieuwenhuizen and D.~V. Vassilevich, {\em Class. Quant. Grav.} {\bf 22},
  5029 (2005).

\bibitem{Grumiller:2006xz}
D.~Grumiller and R.~Jackiw, {\em Phys. Lett.} {\bf B642}, 530 (2006).

\end{thebibliography}

\end{document}